\documentstyle[buckow]{article}
\begin {document}
\def\titleline{%
3-dimensional low-energy topological invariants}
%\newtitleline
\def\authors{
Ma{\l}gorzata Bakalarska and Bogus{\l}aw Broda 
}
\def\addresses{
Department of Theoretical Physics, University of
{\L}\'od\'z\\
Pomorska 149/153, PL-90-236 {\L}\'od\'z, Poland\\
gosiabak@krysia.uni.lodz.pl\\
bobroda@krysia.uni.lodz.pl
}
\def\abstracttext{
A description of the one-loop approximation 
formula for the partition function of a three-dimensional abelian version 
of the Donaldson-Witten theory is proposed. The one-loop
expression is shown to contain such topological invariants of a three-dimensional manifold ${\cal M}$ like the Reidemeister-Ray-Singer torsion $\tau_R$ and Betti numbers.
}
\large
\makefront
{\bf 1.} In three dimensions (3d), traditionally, we have the two important well-known topological 
quantum field theories (of ``cohomological'' type): non-abelian
$SU(2)$ topological gauge theory of flat connection and 3d 
version of the topological Seiberg-Witten theory. 
The former is a 3d twisted ${\cal N}=4$ SUSY $SU(2)$ pure gauge
theory or a 3d version of the Donaldson-Witten theory, 
and ``by definition'' it describes the Casson invariant, which 
appropriately (i.e. with sign) counts the number of flat $SU(2)$ connections \cite{1}. 
The latter is a dimensional reduction of the topological Seiberg-Witten 
theory or a twisted version of 3d ${\cal N} = 4$ SUSY $U(1)$ gauge theory with 
a matter hypermultiplet \cite{2}, \cite{3}. Also this theory describes an interesting  non-trivial topological invariant of 
3d manifolds pertaining to 3d Donaldson-Witten invariant which has been conjectured 
to be equivalent to already known topological invariants and in particular to topological (Reidemeister) torsion \cite{4}. The conjectures are
physically  strongly motivated by the fact that the both theories 
can be derived, performing dimensional reduction and/or taking low-energy limit, from 4d ${\cal N}=2$ SUSY $SU(2)$ pure gauge 
theory \cite{5} corresponding via twist to Donaldson-Witten theory. Interestingly, 
it follows from \cite{6} that low-energy effective theory for
both 3d ${\cal N}=4$ SUSY $SU(2)$ pure gauge case as well as for the ${\cal N}=4$ SUSY abelian one with matter hypermultiplet is an abelian pure 
gauge theory. Therefore, we should essentially deal with 3d twisted version of the abelian theory---topological
gauge theory of flat $U(1)$-connection \cite{7}.
In dual variables, we have a 3d ${\cal N}=4$ SUSY $\sigma$-model with a hyper-k\"ahler manifold as a target space. Twisted version of this model has been analyzed in detail by Rozansky and Witten in \cite{8}. Here, we will consider a corresponding topological model in original (non-dual) variables. In fact, we will confine ourselves to the lowest-order one-loop calculation which nevertheless gives a (exact) topological invariant---Reidemeister-Ray-Singer torsion---as it should. In spite of a direct interpretation of our model as a low-energy version of the above-mentioned gauge models, we can treat it as a stand-alone one as well. Furthermore, we can even treat its (one-loop) gaussian approximation as an independent model, as we actually do in this note.

{\bf 2.} Topological action of our theory assumes the following form
\begin{eqnarray}
S_{\rm eucl} &=& {\int}_{\cal M} d^3 x \sqrt{g} \left\{ \frac{1}{4}
            F^{mn}F_{mn} - \frac{1}{2} \varphi {\nabla}^m 
            {\nabla}_m \varphi - \bar{B} {\nabla}^m {\nabla}_m B
             - \frac{1}{\sqrt{g}}{\varepsilon}^{kmn} {\chi}_k {\partial}_m {\psi}_n
             + \mbox{}\right.\nonumber\\  
             &&\left.\mbox{} - \omega {\nabla}^m {\chi}_m +
             \rho {\nabla}^m {\psi}_m \right\},
  \label{1}
\end{eqnarray}
where $F_{mn}$ is the usual $U(1)$-gauge field strength, $B$ is a complex scalar,
$\varphi$ is a real scalar, and the fields $\omega$, $\rho$, ${\chi}_k$,
${\psi}_k$ are fermionic fields. The covariant derivative of fermions
and bosons, here denoted as ${\nabla}_m$, is defined using the
Levi-Civita connection on ${\cal M}$ with a metric $g_{mn}$, where
$m, n, k = 1,2,3$.

Gauge invariance of the action (\ref{1}) requires gauge 
fixing and introduction of the Faddeev-Popov ghost fields 
$c$, $\bar{c}$
\begin{equation}
  {\cal L}_{\rm gauge}= \frac{1}{2}({\nabla}^m {A}_m)^2 - \bar{c} {\nabla}^m 
                {\nabla}_m c.
  \label{2}     
\end{equation}
Supplementing the action (\ref{1}) with  (\ref{2}), 
we obtain
\begin{eqnarray}
S_{\rm eucl} &=& {\int}_{\cal M} d^3 x \sqrt{g} \left\{ \frac{1}{4}
             F^{mn} F_{mn} - \frac{1}{2} \varphi {\nabla}^m
             {\nabla}_m \varphi - \bar{B} {\nabla}^m {\nabla}_m B
             + \mbox{}\right.\nonumber\\
             &&\mbox{} -  \frac{1}{\sqrt{g}}{\varepsilon}^{kmn} {\chi}_k {\partial}_m
             {\psi}_n -  \omega {\nabla}^m {\chi}_m +\rho {\nabla}^m
             {\psi}_m + \mbox{} \nonumber \\
             &&\left.\mbox{} + \frac{1}{2}({\nabla}^m {A}_m)^2 -
             \bar{c}{\nabla}^m {\nabla}_m c \right\}.
  \label{3}
\end{eqnarray}
It will appear that the partition function $Z$ corresponding to (\ref{3})  essentially consists of the Reidemeister-Ray-Singer torsion of 3d manifold $\cal M$.
Namely,
\begin{equation}
        Z({\cal M}) = \int [{\cal D}X] {\rm exp} \left\{ - S_{\rm eucl}[X]\right\},
  \label{4}
\end{equation}
where the integration measure $[{\cal D}X]$ is taken over all the
fields 
\[
 (A_m, \varphi, B, \bar{B}, \omega, \rho, {\chi}_k, 
{\psi}_k, c, \bar{c}).
\]
The path integral over the bosonic field $\varphi$ gives the factor 
\begin{equation}
      ({\rm det}^{\prime} (-{\triangle}_0))^{-\frac{1}{2}},
\end{equation}
where ${\triangle}_0 = {\nabla}^m {\nabla}_m$ is a Laplacian
acting on zero-forms on $\cal M$ and ${\rm det}^{\prime}$ means that we
exclude zero modes.
The integration over the usual ghosts $c$ and $\bar{c}$ gives 
${\rm det}^{\prime}(- {\triangle}_0)$, while the $B$- and $\bar{B}$-%
integrations give its inverse $({\rm det}^{\prime}(-{\triangle}_0))
^{-1}$. So, these modes are canceled.
Now let us introduce an operator $L_{-}$ which acts on the
direct sum of zero- and one-forms on ${\cal M}$ \cite{8}
\begin{equation}
L_{-}(\rho, {\chi}_k) = (-{\nabla}^m {\chi}_m, {\nabla}^k \rho 
                        +\frac{1}{\sqrt{g}}{\varepsilon}^{kmn} {\partial}_m {\chi}_n),
 \label{5}
\end{equation} 
and similarly for $(\omega, {\psi}_k)$.

Following \cite{8} we define  a scalar product
\begin{equation}
     \left<\rho, {\chi}_k  \mid  {\rho}^{\prime}, {\chi}_k^{\prime}\right> 
        = {\int}_{\cal M} d^3 x \sqrt{g} (\rho {\rho}^{\prime} +
          {\chi}^k {\chi}_k^{\prime}).
\end{equation}
Then the part of the action (\ref{3}) containing $\omega$,
$\rho$, ${\chi}_k$ and ${\psi}_k$ becomes a quadratic form 
\begin{equation}
   \frac{1}{2}\left<\omega, {\psi}_k \mid L_{-} \mid \rho, {\chi}_k \right>  
 -\frac{1}{2}\left<\rho, {\chi}_k \mid L_{-} \mid \omega, {\psi}_k \right>.  
\end{equation}

The fermionic one-loop contribution \cite{9}, with zero modes removed, is
\begin{equation}
         {\rm det}^{\prime} L_{-}.
\end{equation}

Finally, the integration over the gauge field ${A}_m$ gives 
\begin{equation}
         ({\rm det}^{\prime} ( - {\triangle}_{1}))^{-\frac{1}{2}},
\end{equation}
so that the total one-loop contribution of non-zero modes is
\begin{equation}
         \frac{{\rm det}^{\prime} L_{-}}{\left({\rm det}^{\prime}
         ( - {\triangle}_{0})\right)^{\frac{1}{2}}\left({\rm det}^{\prime}
         ( - {\triangle}_{1})\right)^{\frac{1}{2}}}.
  \label{6}
\end{equation}

The absolute value of the ratio of the determinants in
eq.(\ref{6}) is related to the Reidemeister-Ray-Singer  analytic torsion ${\tau}_{R}({\cal M})$ \cite{8}, \cite{10},
\cite{11}.
The formula for the absolute value of the ratio of the determinants is 
\begin{equation}
\left | \frac{{\rm det}^{\prime} L_{-}}{\left({\rm det}^{\prime}
    (- {\triangle}_{0})\right)^{\frac{1}{2}} \left({\rm det}^{\prime}
    (- {\triangle}_{1})\right)^{\frac{1}{2}}} 
    \right | = {\tau}_{R}^{-2}({\cal M}). 
 \label{7}
\end{equation}

{\bf 3.}
It is not the whole story, because the partition function is, in principle, plagued by zero modes, which we have removed by hand. There are the following four sorts of zero modes, which should be treated differently (the number of some of them may depend on topology of $\cal M$): (1) 3 scalar boson zero modes, corresponding to $\varphi$, $B$ and $\bar{B}$; (2) $b_1$ ($=1^{\rm st}$ Betti number of $\cal M$) vector boson zero modes of $A_m$; (3) 1 ghost zero mode for $c$, $\bar{c}$; (4) 2 scalar fermion zero modes of $\omega$, $\rho$, and $2b_1$ one-form zero modes for $\chi_k$, $\psi_k$.
The ghost zero mode can be removed instantaneously because it should not be present in the partition function from the very beginning at all, as the gauge transformation corresponding to the constant (zero) mode acts trivially on $A_m$. Possibly, we could integrate out eq.(\ref{4}) with respect to vector zero modes because it corresponds to integrating over $b_1$-torus. The rest of boson and fermion zero modes produces infinities and zeros in the partition function respectively. In this gaussian approximation/version there is no other possibility than to remove them by hand. They can be properly handled only in higher orders. We need a compact-space valued scalar field ($\sigma$-model) to integrate out scalars, and higher-order fermion terms to saturate fermion zero modes. As usual, boson and ghost zero modes yield appropriate powers in the Planck constant in front of eq.(\ref{4}).

\bigskip
The second author (B.~B.) would like to thank the organizers for their kind invitation and assistance during the symposium. The work has been supported by the KBN grant 2~P03B~084~15.

%%%%%%%%%%%%%%%%%%%%%%
%%%%%%%%%%%%%%%%%%%%%%

\end{document}